\documentstyle[prl,multicol,aps,epsf]{revtex}
\begin{document}
%\begin{flushright}
%{\bf ITP-UU-00/37}
%\end{flushright}
%\begin{center}
%{\large\bf Discrete symmetry in spinor Bose-Einstein Condensates}\\[2ex]
%Fei Zhou\\[1ex]
%{\it Physics Department, Princeton university, Princeton, NJ 
%08544, USA}\\
%{\it ITP, Utrecht University, Princetonplein 5,
%3584 CC Utrecht, The Netherlands\footnote{Permanent address}} 
%\end{center}

\title{Spin Correlation and Discrete Symmetry in Spinor Bose-Einstein 
Condensates}
\author{Fei Zhou}
\address{Physics Department, Princeton university, Princeton, NJ 
08544, USA}
\address{ITP, Utrecht University, Princetonplein 5,
3584 CC Utrecht, The Netherlands\footnote{Permanent address}} 
\maketitle

\begin{abstract}
We study spin correlations in Bose-Einstein
condensates of spin
1 bosons with scatterings dominated by a total spin equal
$2$ channel. 
We show that the low energy spin dynamics in the system can be mapped into 
an $o(n)$ nonlinear sigma model(NL$\sigma$M).
$n=3$ at the zero magnetic field limit and $n=2$ in the presence
of weak magnetic fields.
In an ordered phase,
the ground state has a discrete $Z_2$ symmetry and the degeneracy
space is $[U(1)\times
S^{n-1}]/Z_2$.
We explore consequences of the discrete symmetry and
propose some measurements to probe it.

\end{abstract}

\pacs{74.50. +r, 05.20. -y, 82.20. -w}

\begin{multicols}{2}

%\narrowtext

Recently, there has been a burst of theoretical and experimental
activities on spinor
Bose-Einstein condensates\cite{Myatt,Stamper,Stenger,Ho,Ohmi,Law}. 
An optical trap confines alkali atoms independent of spins
and liberates spin degree of freedoms\cite{Stamper,Stenger}. 
For sodium($^{23}Na$) or rubidium($^{87}Rb$) atoms with nuclear spin
$I=3/2$ and electrons at $s$ orbits, 
the energy splitting 
between hyperfine multiplets 
is of order $100 mk$.
At temperatures as low as $100nk$, $^{23}Na$
and $^{87}Rb$ atoms can be considered as simple bosons with a hyperfine
spin $F=1$. The ground state of N spin $1$ noninteracting bosons  has
$(N+1)N/2$ folds spin degeneracy, by contrast to its magnetically trapped 
cousins. Presumably, hyperfine spin-dependent scattering lifts
the degeneracy and leads to a spin correlated state. 
Optically trapped BEC therefore sets up a platform
for studying quantum magnetism in many boson systems and
adds a new dimension to already extremely rich physics in these 
systems.

The spin-dependent two
-body interaction in BEC is characterized by
$U_2({\bf r_1}, {\bf r_2})=\delta({\bf r}_1-{\bf r}_2)[c_0 + 
c_2 {\bf F_1} \dot {\bf F_2}]$, as suggested in an early paper\cite{Ho}. 
Here $c_0=(g_0+2g_2)/3$, $c_2=(g_2 -g_0)/3$;
$g_F=4\pi \hbar^2 a_F/M$, M is the mass of the atom and
$a_F$ is the s-wave scattering length in the total spin $F$ channel.  
Thus,
the spin correlation in a BEC is determined by $c_2$. For
$^{87}Rb$, $g_2 < g_0$ or
$c_2 <0$
and the scattering
is dominated by the total spin $F=0$ channel.
In the ground state, all spins of atoms prefer to 
align in a certain direction and have a maxima magnetization\cite{Ho}.

For $^{23}Na$ studied experimentally\cite{Stamper}, 
the scattering between $^{23}Na$ atoms is dominated by 
the total spin $F=2$ channel, i.e.
$g_2 > g_0$.
The scattering between $^{23}Na$ atoms thus leads to an
"antiferromagnetic" spin correlation.
Efforts have been made to understand
the ground state properties, exact excitation spectra, collective modes
\cite{Ho,Ohmi,Law}. 
Many interesting predictions, such as spin waves,
spin mixing dynamics were made
for $^{23}Na$ BEC where $c_2 >0$.

In this paper, 
we show the spin dynamics in BEC with $c_2 >0$ is
characterized by an $o(n)$ nonlinear sigma model (NL$\sigma M$) of $n$
components. $n=3$ at zero magnetic field limit and
$n=2$ in the presence of a weak magnetic field.
Spin correlations in spinor BEC can be studied in the context of
the NL$\sigma$M.
We identify that the internal order parameter space for
BEC as $[S^1\times S^2]/Z_2$(zero field limit) and the ground state
is nematically ordered. 
We explore consequences of the discrete $Z_2$ symmetry.

To describe the spin correlated BEC, 
it is most convenient to introduce 
Weyl representation of $SU(2)$ involving polynomials of a unit vector $(u,
v)$\cite{Arovas}. 
Each unit vector is represented by a point ${\Omega}$ on a sphere
with polar coordinates $(\theta, \phi)$; namely 
$u=\exp(i\phi/2)
\cos(\theta/2)$, $v=\exp(-i\phi/2)\sin\theta/2$.
The corresponding hyperfine spin operators are 
$F^+=u\partial/\partial v$,
$F^-=v\partial/\partial u$,
and $F_z=(u\partial /\partial u -v \partial /\partial v)/2$.
The scalar product between two wavefunctions $g$ and $f$ is 
defined as $\int g^*(u,v)f(u, v)d{\Omega}/4\pi$.
(We reserve ${\bf \Omega}$ for the spin rotations discussed below.)

Under spin rotations ${\cal R}=\exp(iF_z\chi_1/2)
\exp(iF_y \theta_1/2)\\
\exp(iF_z\phi_1/2)$, $u$ and $v$ transform into

\begin{eqnarray}
&&u({\bf \Omega_1},\chi_1)=  \exp(i{\chi_1}/{2}) \nonumber \\
&&(\cos{\theta_1}/{2} \exp(-i{\phi_1}/{2})u + \sin 
{\theta_1}/{2}\exp(i{\phi_1}/{2})v), \nonumber \\ 
&& v({\bf \Omega_2},\chi_2)=\exp(-i{\chi_2}/{2})
\nonumber \\
&&(-\sin{\theta_2}/{2}
\exp(-i{\phi_2}/{2})u+\cos{\theta_2}/{2}\exp(i{\phi_2}/{2})v),
\end{eqnarray}
where ${\bf \Omega}_{1,2}=(\theta_{1,2}, \phi_{1,2})$.
Spin-1 wavefunctions are polynomials of degree 2  in $u$ and $v$.
$\sqrt{3} u^2$, $\sqrt{6} uv$, $\sqrt{3} v^2$ correspond to
$m=1, 0, -1$ states. 
All $F=1$ states can also be expressed in term of $\sqrt{6}u({\bf
\Omega_1})v({\bf \Omega_2})$ with ${\bf \Omega}_{1,2}$ properly chosen.

The Hamiltonian for spin-1 bosons can be written as

\begin{eqnarray}
&&{\cal H}=
-\frac{1}{2M}
\sum_\alpha
\nabla^2_\alpha 
+\sum_{\alpha,\beta}
[\frac{c_0}{2} + \frac{c_2}{2}{\bf F}_\alpha \cdot {\bf F}_\beta] 
\delta({\bf r}-{\bf r}')
\nonumber \\
&&- \sum_\alpha {\bf F}_{z\alpha}g \mu_B H. 
\end{eqnarray}
The second term is hyperfine spin-dependent interaction with $c_2 >0$ and
the last term is the coupling with an 
external magnetic field ${\bf H}=H{\bf e}_z$;
$g$ is a Lande factor of an atom and $\mu_B$ is the Bohr magneton.

The wavefunction of N spin-1 Bosons generally can be written 
as\cite{Zhou}

\begin{equation}
\Psi(\{{\bf r}_\alpha\})={\cal P}\Pi_{\alpha=1...N}
\Phi_{N_\alpha}({\bf r}_\alpha)
\sqrt{6}u_\alpha({\bf \Omega}_{1\alpha}({\bf r}_\alpha))
v_\alpha({\bf \Omega}_{2\alpha}({\bf r}_\alpha)).
\end{equation}  
${\cal P}$ is the permutation of $\{N_\alpha \}$,
$\{ {\bf \Omega}_{1\alpha}, {\bf \Omega}_{2\alpha} \}$. 
$N_\alpha$ labels an one-particle orbital state. Phases 
$\chi_{1,2}$ in Eq.1 are absorbed by a gauge transformation of the 
wavefunction $\Phi_{N_\alpha}$ introduced above. 
For BEC under consideration, we
take ${\bf \Omega}_{1\alpha,2\alpha}={\bf \Omega}_{1,2}({\bf r})$ and
$\Phi_{N_\alpha}({\bf r})=\Phi({\bf r})$
($\Phi({\bf r})$ is a complex scalar field).
By introducing ${\bf n}({\bf r})=({\bf \Omega}_1 + {\bf \Omega}_2)/2$,  
${\bf L}({\bf r})=({\bf \Omega}_1-{\bf \Omega}_2)/2$ and 
$\Phi({\bf r})=\sqrt{\rho({\bf r})}
\exp(i\chi({\bf r}))$, we derive from Eqs.2,3 an effective Hamiltonian as a 
function of
two pairs of variables : $\{ {\bf n}(\bf r), {\bf L}({\bf r}) \}$
and $\{ \rho ({\bf r}), \chi({\bf r}) \}$.
These are collective variables of the
N interacting spin-1 Bosons, which describe the spin dynamics and 
phase dynamics respectively.
In this representation, the hyperfine spin dependent interaction in Eq.2
is mapped into ${\cal H}_s=\int d{\bf r}c_2 {\bf L}^2({\bf r}) \rho^2({\bf 
r})$, which only depends on collective variable ${\bf L}$ when $\rho({\bf 
r})$ is taken as a constant.
This indicates that each atom acquires an inertial $I_0=1/2c_2 \rho$ in 
the presence
of hyperfine spin dependent scatterings with $c_2 >0$;
the rotation energy in the presence of a 
finite spin moment
${\bf L}$ is thus ${\bf L}^2/2 I_0$.

In the most interesting limit, we can introduce
the local spin density as
${\bf l}({\bf r})={\bf L}({\bf r})\rho({\bf r})$.
${\bf n}$ and ${\bf l}$ satisfy the constrain
${\bf n}({\bf r}) \cdot {\bf  l}({\bf r})=0$.
Commutation relations between
$\rho$ and $\chi$,  ${\bf n}({\bf r})$ and ${\bf l}({\bf r})$ are
given as
$[\rho({\bf r}),\chi({\bf r'})]=i\hbar \delta({\bf r}-{\bf r'})$;
$[{\bf n}_\alpha({\bf r}), {\bf n}_\beta({\bf r}')]=0$,
$[{\bf l}_\alpha({\bf r}), {\bf n}_\beta({\bf r}')]=i\hbar
\epsilon^{\alpha\beta\gamma}{\bf n}_\gamma \delta({\bf r}-{\bf r}')$, 
$[{\bf l}_\alpha({\bf r}), {\bf l}_\beta({\bf r}')]=i\hbar
\epsilon^{\alpha\beta\gamma}{\bf l}_\gamma \delta({\bf r}-{\bf r}')$. 
$\epsilon^{\alpha\beta\gamma}$ is an antisymmetric tensor.
These identities are valid 
when ${\bf L}$ per atom is much less than unity
and ${\bf n}({\bf r})$ can be considered as a classical "vector",
components of which commutate with each other.
The corresponding Lagrangian density can be derived as 
${\cal L}={\cal L}_{s} +{\cal L}_c +{\cal L}_{sc}$\cite{Zhou},
with

\begin{eqnarray}
&&{\cal L}_c\approx \frac{\rho}{2M} [
(\nabla {\chi}({\bf r}))^2
+\frac{1}{v_c^2}(\partial_{\tau} {\chi})^2],
\nonumber \\
&&{\cal L}_s=\frac{\rho}{2M} [
(\nabla {\bf n}({\bf r}))^2
+\frac{1}{v_s^2}(\partial_{\tau} {\bf n})^2],
\nonumber \\
&&{\cal L}_{cs}=\frac{\rho}{M} \frac{1}
{v^2_s} [{\bf n}\times \partial_\tau {\bf n}
\cdot  {\bf n} \times ({\nabla \chi} \cdot \nabla){\bf n}].
\end{eqnarray}
Here $\rho=\rho(0)$, 
$v_c=\sqrt{2\rho c_0/M}$,
$v_s=\sqrt{{2\rho c_2}/M}$. 
We introduce $\tau=it$ as the imaginary time. 
Nonlinearality is imposed via a constraint
$|{\bf n}^2|=1$ at a low frequency limit.
Eq.4 is the main result of the mapping and
we have kept contributions which are lowest order in terms of 
$\partial_\tau$
and $\nabla$. 
${\cal L}_c$ is taken in a Gaussian
approximation and should be replaced by a
full Gross-Pitaveskii Lagrangian in general. We will be mostly 
interested in the spin sector and 
simplification in ${\cal L}_c$ doesn't affect conclusions here.
${\cal L}_s$ in Eq.4 represents an $o(3)$ NL$\sigma$M.

The last term ${\cal L}_{sc}$ characterizes
a coupling between a spin rotation and the superflow in BEC
due to Berry's phase.
This term, however, is linear in $\partial_\tau$
and quadratic in spatial
gradient and is negligible compared with ${\cal L}_s, {\cal L}_c$ at
the long wave length limit. Particularly,
such a coupling vanishes in a configuration where ${\bf L}$ is zero.

At the zero field limit, there exists 
a saddle point solution for the spin sector
${\bf n}({\bf r})={\bf n}_0$, 
${\bf l}({\bf r})=0$. ${\bf n}_0$ lives on a unit sphere.
By expanding Eq.4 around the saddle point solution,
we obtain spin waves with sound like spectrum $\omega=
\sqrt{4c_2\rho/M}k$, which can also be obtained in the Gross-Pitaveskii 
approach\cite{Ho}. However, Eq.4 here
is valid for any point ${\bf \Omega}_1\sim {\bf \Omega}_2$ on the 
unit sphere.
The effective NL$\sigma$M derived here allows us to describe
spin correlated states well beyond the
Gross-Pitaveskii approach.

For a symmetry broken state,
the ground state wavefunction in Eq.3 with 
${\bf \Omega}_1 ={\bf \Omega}_2 ={\bf n}$ 
is invariant under a global transformation ${\bf n}, \chi \rightarrow 
-{\bf n}, \chi+\pi$;
\begin{equation}
\Psi({\bf n}, \chi)=\Psi(-{\bf n}, \chi+\pi),
\Psi({\bf n})=(-1)^N \Psi(-{\bf n}) 
\end{equation}
where $\chi$ is the phase of the scalar field $\Phi(x)$ introduced before.
In obtaining this symmetry, we notice $u({\bf n})=\exp(i\pi/2)
v(-{\bf n})$, with $\pi/2$ from a phase of a spin-1/2 particle
under a $180^0$ rotation.
Eq.5 shows that the many-body wavefunctions characterized
by $({\bf n},\chi)$ and $(-{\bf n}, \chi +\pi)$ are indistinguishable. 
Thus, the spinor BEC under considerations 
has a discrete $Z_2$ symmetry and the order parameter space is
${\cal R}=[S^1\times S^2]/Z_2$,
with $Z_2$ as a two-element group of integers modulo 2.

As in a classical uniaxial nematic liquid crystal where 
diatomic molecules are indistinguishable upon an inversion of
their directors ${\bf n}$ and the internal
order parameter space is $S^2/Z_2$, the $Z_2$ symmetry identified 
here
also indicates that there exists a tensor order parameter\cite{de 
Gennes}. 
For the purpose of demonstration,
let us introduce a "director" 
${\bf d}_x=[uv^*+vu^*]/{2}$, 
${\bf d}_y=[uv^*-vu^*]/{2i}$,
${\bf d}_z=[uu^*-vv^*]/{2}$. One then can show for the ground state 
wavefunction in Eq. 3,
$<{\bf d}>=0$ but nematic order parameter
$Q_{\alpha\beta}=1/{\rho}[<{\bf d}_\alpha{\bf d}_\beta>- 
{1}/{3}Tr <{\bf d}_\alpha{\bf d}_\beta> ]$
is nonzero. Here $<>$ stands for an average taken over the ground state wave 
function $\Psi$. In fact,

\begin{eqnarray}
&&Q_{\alpha\beta}=\frac{2}{5}[-3{\bf n}_{0\alpha} {\bf 
n}_{0\beta} +\delta_{\alpha\beta}].
\end{eqnarray}
According to Eq.6,
the director ${\bf d}$ aligns in a plane perpendicular to ${\bf n}_0$.

Topological defects in this case are of particular interest because
of the discrete symmetry in Eq.5. 
Following the general theory for the classification of defects in 
a symmetry broken state,
the $Z_2$ symmetry leads to $\pi$ spin disclinations superimposed with 
half vortices, which 
otherwise don't exist. The 
corresponding wavefunction of composite linear singularities
($Z_2$ strings) is

\begin{eqnarray}
&&\lim_{\xi \rightarrow \infty}{\bf n}(\xi)=Re 
(\frac{\xi-\xi_0}{|\xi-\xi_0|})^{m+1/2}
{\bf e}_{x}+ Im(\frac{\xi-\xi_0}{|\xi-\xi_0|})^{m+1/2}{\bf e}_{y},
\nonumber \\
&&\lim_{\xi\rightarrow \infty}{\bf v}_s(\xi)=
\frac{1/2+n}{M|\xi-\xi_0|}
\nonumber \\&&
[Im(\frac{\xi-\xi_0}{|\xi-\xi_0|}){\bf e}_x -
Re(\frac{\xi-\xi_0}{|\xi-\xi_0|}){\bf e}_y].
\end{eqnarray}
Here $\xi=x+iy$
and lines are located at 
$\xi_0=x_0+iy_0$; $n,m$ are integers and 
${\bf v}_{s}$ is superfluid velocity\cite{Volovik}.
Each string is characterized by $(m,n)$. However, $(-1,n)$, $(\pm 3, n)$,
$(\pm 5, n)$ strings can be obtained by deforming string $(1,n)$ and 
are homotopical identical to $(1,n)$.

In a composite string given in Eq.7,
${\bf n}$ evolves into $-{\bf n}$ along a loop 
enclosing $\xi_0$. 
The corresponding spin wavefunction changes its sign under an inversion 
${\bf n} \rightarrow  -{\bf n}$, following the 
identity $u({\bf n})v({\bf n})=-u(-{\bf n})v(-{\bf n})$.
A superflow ${\bf v}_s$ of a half vortex is present to
compensate the $\pi$ phase under ${\bf n}\rightarrow -{\bf n}$
rotation and ensure the single valuedness of the 
wavefunction. 
This composite structure is different from a linear defect in a classical
nematic liquid where $\pi$ disclinations are free topological 
excitations. In fact, in a coherent spinor BEC, a bare $\pi$-spin 
disclination carries a cut 
along which phase changes abruptly from $\pi$ to $2\pi$ or $0$. 
This cut starting at the disclination ends only at the boundary of the 
system and costs an energy linear in 
term of the size of BEC. 
For a similar reason,
the energy cost to have a $\pi$ disclination
and a half-vortex separated at a distance $L$ is linearly proportional
to $L$. 
Composite strings in Eq.7 should be considered as
results of confinement of
$\pi$ spin disclinations and half vortices 
in spinor BEC\cite{Duncan}.

In the presence of an external magnetic field along $z$ direction,
${\bf \Omega}_{1,2}={\bf n}\pm {\bf e}_z g\mu_B H/4c_2 \rho$ and
${\bf n}$ satisfies constrain

\begin{eqnarray}
{\bf n} \cdot \frac{\mu_B H}{4c_2 \rho} {\bf e}_z=0.
\end{eqnarray}
Obviously,
an external magnetic field breaks the $S^2$ symmetry and confines 
low frequency sector of ${\bf n}$ in
a plane perpendicular to ${\bf H}$ itself.
The Lagrangian in the presence of a magnetic field
is that of an $O(2)$ NL$\sigma$M; it has
$S^1$-symmetry at the frequency $\omega \ll \mu_B H$. 
At the high frequency limit,
${\bf n}$ precesses in a field
$4c_2 {\bf l}(x)$, much larger than the external field and 
$S^2$-symmetry is restored.

As a consequence,
the  order parameter space for the quantum spin nematic state 
in an external magnetic field 
is ${\cal R}=[S^1\times S^1]/Z_2$.
The nematic order parameter 
$Q_{\alpha\beta}$ is still given by Eq.7, with the easy
plane of ${\bf d}$ parallel to the external field.
Wavefunctions for linear defects are
of same forms as those in Eq.7, but with all strings $(m,n)$
homotopically distinguishable.

We have restricted ourselves to the weak magnetic field limit
and neglected the possible quadratic Zeemann shift
${\cal H}_{QZ}=\sum_\alpha Q H^2 F^2_{z\alpha}$ 
(the external field $H$ is along ${\bf e}_z$ direction).
Inclusion of the quadratic Zeemann shift yields an additional 
term ${\cal L}_{QZ}=\int \rho Q H^2 ({\bf n}^2_x +{\bf n}^2_y) dx$
to the NL$\sigma$M derived.
The main effect of this contribution is to align ${\bf n}$ along
the external field. When this shift dominates,
the ground state is left with a double degeneracy:
${\bf n}={\bf e}_z$ and ${\bf n}=-{\bf e}_z$.
The spin wave develops an energy
gap of order $QH^2$.
We will focus on the zero magnetic field case
in the rest of discussions.

In general,
following Eq.4,
spin correlated BEC can be studied 
by considering a NL$\sigma$M, 

\begin{eqnarray}
&&{\cal L}_s=\frac{1}{2f}(\partial_\mu {\bf n})^2, {\bf n}^2=1;
\nonumber \\
&& f=(16\pi)^{1/2}(\rho
\Delta a^3)^{1/6}, \Delta a=\frac{a_2-a_0}{3}.
\end{eqnarray} 
We introduce dimensionless length and time:
$\tilde{{\bf r}}={\bf r}\rho^{1/3}$, $\tilde{\tau}=\tau v_s\rho^{1/3}$.
Derivatives $\partial_\mu$ are defined as
$(\partial_{\tilde{\tau}},\partial_{\tilde x},\partial_{\tilde y},
\partial_{\tilde z})$.
$f^{-1}$ is a square root of the ratio between potential energy
at an interatomic scale
$\hbar^2 \rho^{2/3}/2 m$ and zero point kinetic(rotation) energy
$c_2\rho/2$ of an individual atom.
The zero point motion is absent for noninteracting bosons but
gets stronger when $c_2$ increases, or the effective inertial $I_0$
gets smaller. 
The $o(3)$ NL$\sigma$M has order and disordered
phases 
at $d > 1$,
depending on the parameter $f$.
Most of qualitative results about spin correlations in BEC can be obtained 
in a renormalization group (RG) approach\cite{Zhou}.
For $\rho \Delta a^3=10^{-6}$ as is in experiments, a
long range nematic order should be observed.

In $1D$ case, at zero temperature and zero field, 
one should expect there will be no long range order 
and the state is nematically
disordered, following the RG
results of NL$\sigma$M\cite{Polyakov}. 
These nematic disordered states mimic the quantum spin 
liquid states proposed in the literature of Heisenberg antiferromagnetic 
systems(HAFS). 
However, we notice that
the NL$\sigma$M derived from the microscopic Hamiltonian in this paper
doesn't have a $\theta$-term
${\cal L}_\theta={\theta}/{4\pi} \int d\tau dx {\bf n} \cdot 
{\partial {\bf 
n}}/{\partial \tau} \times {\partial {\bf n}}/{\partial x}$,
which is generally present in HAFS studied before\cite{Haldane}.  
Absence of a $\theta$-term,
which ensures an energy gap in the excitation spectrum 
of nematic state,
follows a fact that the
Berry's phase under rotations of ${\bf n}$  
vanishes identically\cite{Zhou}.

Two remarks are in order.	
1) At a high density limit, one should also take into 
three-body, four-body elastic scatterings. This can further
modify the short distance dynamics but will not
affect conclusions arrived above in a qualitative way.
2)One should be cautious about the definition of "phase" since
the alkali atoms under investigation are in a long lived metal stable 
gaseous state.
The life time of alkali atomic gas is limited by three-body inelastic 
collisions\cite{Ketterle}. The collision rate is proportional to the 
square of the number density of atoms and increases 
dramatically as the density increases.
This sets a practical limit in order for the quantum disordered 
nematic
liquid to be probed in BEC.

Since the experiment\cite{Stamper} was done in a BEC cloud with a few 
millions sodium atoms, it is also 
particularly interesting to consider the symmetry restoring due to a
finite size effect.
We take a weakly interacting limit where 
quantum fluctuations of finite wave length are negligible.
In a zero mode approximation, 
Hamiltonian becomes,

\begin{equation}
{\cal H}_{z.m}=\rho c_2 \frac{{\bf L}^2}{2N}.
\end{equation}
${\bf L}$ is the total spin of
the N-Bosons system and can be considered as an angular
momentum operator defined on a unit sphere of ${\bf n}$\cite{Zhou}. 
And $[{\bf n}, {\bf n}]=0$, 
$[{\bf L}_\alpha, {\bf n}_\beta]=i\hbar\epsilon^{\alpha\beta\gamma}
{\bf n}_\gamma$, 
$[{\bf L}_\alpha, {\bf L}_\beta]=i\hbar
\epsilon^{\alpha\beta\gamma}{\bf L}_\gamma$. 
The energy spectrum is given by ${\bf L}^2=l(l+1)$,
which is identical to that obtained in \cite{Law}.
$l=0,2,4,....N$, if $N$ is an even number;
and $l=1,3,5,...$ otherwise. The energy gap between lowest lying
excitations is inversely proportional of $N$ and vanishes at large $N$ 
limit.

Now consider a wave packet with
${\bf n}$ confined within
a region centered at ${\bf n}_0={\bf e}_z$ of 
size $\sqrt{<{\delta^2 \bf n}>_0} \ll 1$ on 
the unit sphere at $t=0$.
Spread $<{\delta^2\bf n}>$ is defined as the expectation
value of $ {\bf n}^2_x+ {\bf n}^2_y$.
In spherical polar coordinates $(\theta,\phi)$ where 
${\bf n}$ is a vector represented by $(\sin\theta\cos\phi, 
\sin\theta\sin\phi, 
\cos\theta)$,
the corresponding wave packet(for an even $N$) can be constructed
as

\begin{eqnarray}
&&\Psi(\theta,\phi,t)=\frac{1}{B}
\sum_{l=2n}
\exp(-\frac{l^2}{4\sigma}-i t\frac{l(l+1)c_2\rho}{2N}) 
Y_{l,0}(\theta,\phi).
\nonumber \\
\end{eqnarray}
Here $B=\sum_l \exp(-{l^2}/{2\sigma})$;
$<\delta^2 {\bf n}>_0={A_0}/{\sigma}$ with
constant $A_0$ estimated to be a constant of unity.
$Y_{l,0}(\theta,\phi)$ are spheric harmonics.
$\sigma \gg 1$ but $\sigma/N$ is vanishingly
small. The energy of this wave packet is $\Delta E=2 A_0\sigma c_2 
\rho/N$.
In the limit $N\rightarrow \infty$, 
an infinitesimal external field will stablize this wave 
packet with respect to the 
rotation invariant state.

The spread of ${\bf n}$ at a time $t$ is
\begin{equation}
<(\delta {\bf n})^2>_t=
<(\delta {\bf n})^2>_0+ {4A_0\sigma} (t\frac{c_2\rho}{N})^2,
\end{equation}
which is valid at $t < \tau_c= \sigma /A_0\Delta E$. At $t >\tau_c$,
the spread is of order unity. 
Therefore, 
$\sqrt{<{\delta^2 \bf n}>_t}$ exceeds
the initial spread $\sqrt{<{\delta^2\bf n}>_0}$ 
at a time of order $1/\Delta E$. 
At a longer time  $\tau_c$,
${\bf n}$ starts to explore the whole
unit sphere $S^2$ and the 
rotation symmetry is restored 
due to spin-dependent scatterings in BEC.

Eq.12 also imposes restrictions on measurements.
A measurement of ${\bf n}$ in $^{23}Na$ BEC discussed here
excites the ground
state to an excited state of energy $\Delta E$, where ${\bf n}$ has a 
finite spread 
on $S^2$ and $\Delta E$ is infinitesimal small in the
thermal dynamical limit. The time scale to restore the broken symmetry 
is determined by the two-body spin dependent scattering lengths and the 
number of atoms in BEC. 
A measurement taken at a time scale longer than 
$\sigma/A_0\Delta E$
should reveal the symmetry restoring because
of zero point motions of ${\bf n}$.
With $N \sim 10^7$ and $c_2 \rho \sim 100Hz 
(500nK)$, the symmetry restoring time is about one day
which is much longer than the life time of the trap. 
However, for a smaller cloud with $10^{3-4}$ atoms, the symmetry restoring
can take place within $10-100$secs, before 
recombination processes take place.

Nematically ordered BEC has very fascinating optical 
properties\cite{Zhou}.
In the presence of spin-orbital couplings, dielectric constant is a tensor  
expressed in terms of
$Q_{\alpha\beta}$. 
This suggests that birefringence takes place in the system as a direct
evidence of the hidden $Z_2$ symmetry. 
Another experiment consequence associated with the broken symmetry
is the enhanced small angle light scattering
due to thermal fluctuations of ${\bf n}$.
For $^{23}Na$, the light scattering amplitude can be
four order of magnitude higher than that in an
isotropic BEC and the nematic BEC is optically turbid.
Finally, the $Z_2$ symmetry also implies that there exists a local 
$Z_2$ 
gauge field in BEC. This was recently considered 
and will be published elsewhere\cite{Zhou2}.

FZ would like to thank
E. Demler, J.Ho for many stimulating discussions.
He is particularly
grateful to F. D. Haldane for his guidance and interest on this subject.
This work was partially supported by ARO under DAAG 
55-98-1-0270.

\end{multicols}


\begin{thebibliography}{99}
\bibitem{Myatt}C. J. Myatt et al., Phys. Rev. Lett.
{\bf 78}, 586(1997).
\bibitem{Stamper}D. M. Stamper-Kurn et al., Phys. Rev. Lett.
{\bf 80}, 2027(1998).
\bibitem{Stenger}J. Stenger et al., Nature {\bf 396},
345(1998).
\bibitem{Ho}T. L. Ho, Phys. Rev. Lett. {\bf 81}, 742
(1998); T. L. Ho and S. K. Yip, Phys. Rev. Lett. {\bf 84}, 4031(2000).
\bibitem{Ohmi}T. Ohmi and K. Machinda, J. Phys. Soc. Jpn. {\bf 67}, 
1822(1998).

\bibitem{Law}C. K. Law et al., Phys. Rev. Lett. {\bf 81}, 5257(1998).
\bibitem{Arovas}D. P. Arovas et al., Phys. Rev. Lett. {\bf 60}, 
531(1988);I. Afflect, J. Phys. Condensed Matter 1, 3047(1989).
\bibitem{Zhou}F. Zhou and F.D. Haldane, Preprint ITP-UU-00/51(2000).
\bibitem{de Gennes}P. G. de Gennes, {\it The physics of liquid crystals},
Oxford University Press(1974).
\bibitem{Volovik}
Half vortices 
were also discussed by V. Leonhart and G. E. Volovik, 
JETP Lett. {\bf 72},46(2000). 
\bibitem{Duncan}The vortex core is ferromagnetic with spins
oriented along the direction of the flux, Haldane, unpublished.
\bibitem{Haldane}F. D. M. Haldane, Phys. Rev. Lett. {\bf 50}, 1153(1983).
\bibitem{Polyakov}A. M. Polyakov
{\em Gauge Fields and Strings}, Hardwood academic publishers(1987).
\bibitem{Ketterle}W. Ketterle, D. S. Durfee and D. M. Stamper-Kurn,
in {\it Bose-Einstein condensation in atomic gases, Proceedings of
the international School of Physics Enrico Fermi, Course CXL}, edited
by M. Inguscio, S. Stringari and C. E. Wieman (IOS Press, Amsterdam, 
1999).
\bibitem{Zhou2}E.Demler and F.Zhou, cond-mat/0104409;
E. Demler, F.Zhou and F.D. Haldane, Preprint 
ITP-UU-01/09(2001). 




\end{thebibliography}
\end{document}